\documentclass[twocolumn, prl, showpacs, floatfix]{revtex4}

\usepackage{amsmath}
\usepackage{graphicx}

\begin{document}

\title{``Burning and sticking'' model for a porous material:
suppression of the topological phase transition due to the
backbone reinforcement effect}
\author{A.~S.~Ioselevich, D.~S.~Lyubshin}
\affiliation{Landau Institute for Theoretical Physics RAS,  117940  Moscow, Russia,\\
Moscow Institute of Physics and Technology, Moscow 141700,
Russia.}
\date{\today}
\begin{abstract}
We introduce and study the ``burning-and-sticking'' (BS) lattice
model for the porous material that involves sticking of emerging
finite clusters to the mainland. In contrast with other
single-cluster models, it does not demonstrate any phase
transition: the backbone exists at arbitrarily low concentrations.
The same is true for hybrid models, where the sticking events
occur with probability $q$: the backbone survives at arbitrarily
low $q$. Disappearance of the phase transition is attributed to
the backbone reinforcement  effect, generic for models with
sticking.  A relation between BS and the cluster-cluster
aggregation is briefly discussed.

\end{abstract}
 \maketitle

Single-cluster models of random porous networks which take into
account physical requirements of mechanical stability on all
stages of the system preparation process are promising candidates
for adequate description of realistic porous materials, such as
porous metals \cite{porous-metals}, gels \cite{gels}, aerogels
\cite{aerogels}, etc. The simplest single-cluster models were
introduced and studied in \cite{srbp,srsp}. In these models finite
clusters arising in the process of gradual destruction of randomly
chosen bonds and/or sites in the system are immediately
``repaired'' by means of regeneration of the critical bond/site
whose destruction has lead to violation of connectivity.  The
model with randomly removed (and sometimes regenerated) bonds is
the self-repairing bond percolation (SRBP) model; the model with
randomly removed sites is the self-repairing site percolation
(SRSP) model. In both models the system consists of a single
infinite cluster at all accessible concentrations. However, the
topological properties of this cluster undergo dramatic changes at
a certain critical concentration: the {\it backbone} of the
infinite cluster disappears and the system occurs in a peculiar
tree-like state with anomalous mechanical and electrical
properties caused by the fractal character of this state. We
remind here that the backbone  is defined as an infinite doubly
connected component of an infinite cluster, or, equivalently, as
an infinite {\it block} on the corresponding graph (for the
definition of the term {\it block}, see, e.g. \cite{harary}).
Physically, the backbone is a current-bearing substructure of the
infinite cluster (see, e.g.,
\cite{stauffer-aharony,bunde-havlin}). Note that the existence of
an infinite cluster without a backbone is a very unusual
phenomenon, which is not observed in standard percolation models.

This finite concentration phase transition is also present in  a
one parameter family of hybrid models (which we call SR(S/B)P).
In this case at each step of the sample manufacturing process, with
probability $1-Q$ a randomly chosen bond is removed (and then restored
if necessary) and, with probability $Q$, a randomly chosen site
together with all adjacent bonds is removed (and then restored if
necessary). The properties of the phase transition are, however,
non-universal within the SR(S/B)P family: for example, the fractal
dimension $D_B$ of the backbone near the threshold depends on the
parameter $Q$ (see \cite{srsp}).

\begin{figure}
\includegraphics[width=0.9\columnwidth]{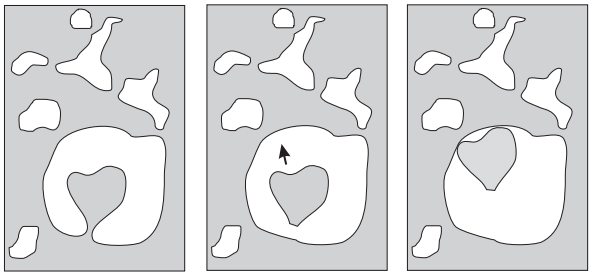}
\caption{``Realistic'' pore forming process. Left panel: just before
a cluster detachment; there is only a thin bridge connecting the
peninsula with the mainland. Middle panel: The bridge is burned, the
island detaches from the mainland and moves toward the wall of
the cavern in the direction of the arrow. Right panel: The
island safely sticks to the wall at two points. }
 \label{stick}
\end{figure}

The site/bond regeneration can only roughly reproduce a realistic
process of manufacturing a porous material. Schematically this
process is as follows: a homogeneous mixture of matrix material
grains and grains of a pore-former (carbon, which can be burned
out, or a soluble polymer) is prepared; then the pore-former
grains are gradually removed. Finite clusters arising in the
process of the pore-former removal immediately fall off and stick
to the surrounding matrix (see Fig.\ref{stick}). It means, in
particular, that the restored bond is not necessarily identical to
the removed one. Moreover, the number of newly created bonds is
larger than one: to establish mechanical stability the cluster
should stick to the matrix at exactly $D$ ($D$ being the space
dimension) points. Would the phase transition be also present for
such realistic process?

\begin{figure}
\includegraphics[width=0.9\columnwidth]{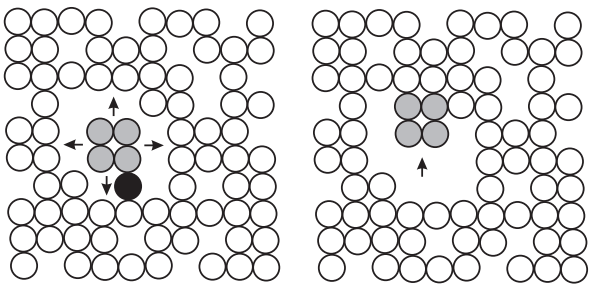}
\caption{The repairing procedure for the ``burning-and-sticking''
model (BS) on a square lattice. The black grain is appointed for
burning at a certain step, and a finite cluster (shown grey) is
ready to be detached (left panel). This cluster is then shifted in
one of four crystallographic directions shown by arrows, and
sticks to the rest of the system (right panel).}
 \label{bs-illustration}
\end{figure}

To answer this question in this paper we introduce another lattice
single-cluster model: the ``burning-and-sticking'' (BS) one. As in
the SRSP model, grains initially occupying all sites of certain
regular lattice are  removed (``burned'') at random, and finite
clusters occasionally created in the burning process are
immediately repaired. However, the repairing procedure which is
launched after detachment of each finite cluster is  different
from the SRSP one. Namely, the disconnected cluster is shifted in
a randomly chosen direction until it sticks to the mainland
(Fig.\ref{bs-illustration}). In the case of a lattice model (which
is only discussed in this paper) the direction of the shift is
chosen from a discrete set of crystallographic axes, so that
upon sticking all grains occupy sites of the same lattice, and the
process can go on.

\begin{figure}
\includegraphics[angle=270,width=0.9\columnwidth]{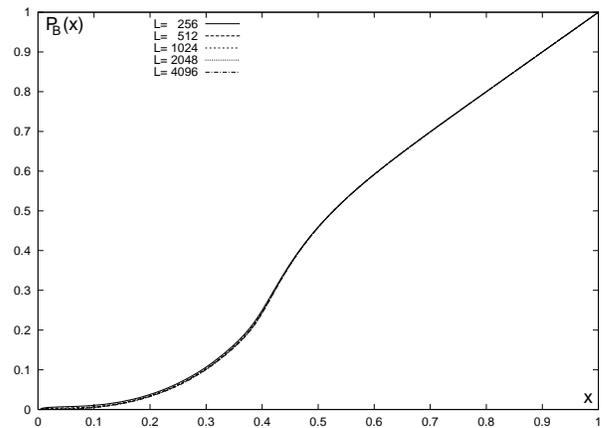}
\caption{Backbone density vs concentration of sites for BS model
with samples of different sizes $L$. The curves almost merge for
$L>512$. There is clearly no phase transition at any finite $x$.
 }\label{bs-backbone}
\end{figure}

In Fig.\ref{bs-backbone} we show the concentration dependence of the backbone
density $P_B(x)$ as obtained from simulations of the BS model on the square
lattice using $L\times L$ samples. The backbone clearly exists
for all $x>0$, and at small $x$ its density vanishes as
\begin{eqnarray}
P_B(x)\propto
 x^{\beta_B}, \label{scaling2}
\end{eqnarray}
with $\beta_B^{\rm (BS)}\approx 2.85(15)$.

\begin{figure}
\includegraphics[width=1\columnwidth]{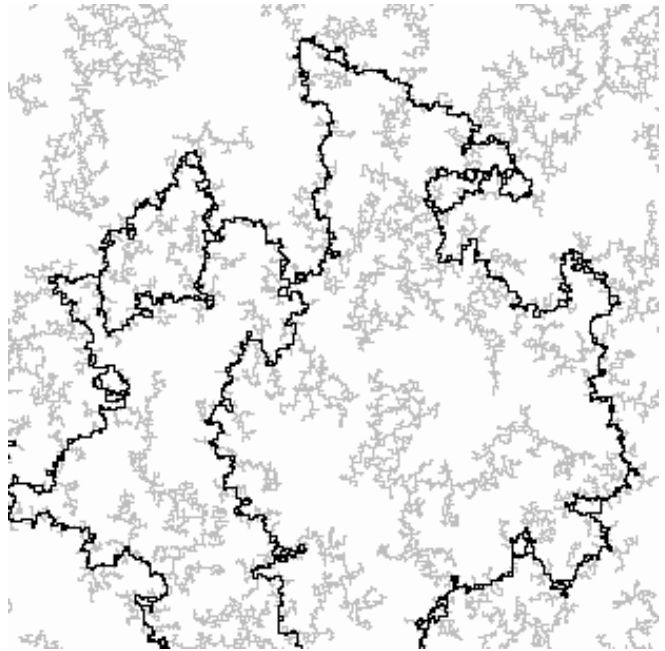}
\caption{ Snapshot of a BS-system at site concentration $x=0.2$.
Backbone has density $0.02742$ and is shown black, dangling ends
are shown grey.  Fragment of size $309\times306$. } \label{bs}
\end{figure}

A fragment of the BS-pattern at relatively low concentration
$x=0.2$ is shown in Fig.\ref{bs}. One notices that the density of
the entire cluster looks much more homogeneous than that of the
backbone. Hence, one can expect that the correlation radius for
total density $\xi_D(x)$ diverges slower than that for the
backbone $\xi_B(x)$. The presence of two different scales
$\xi_D(x)\ll \xi_B(x)$ presumably should make the low-density
behavior of the BS model very rich. This issue is under study now.

\begin{figure}
\includegraphics[width=1\columnwidth]{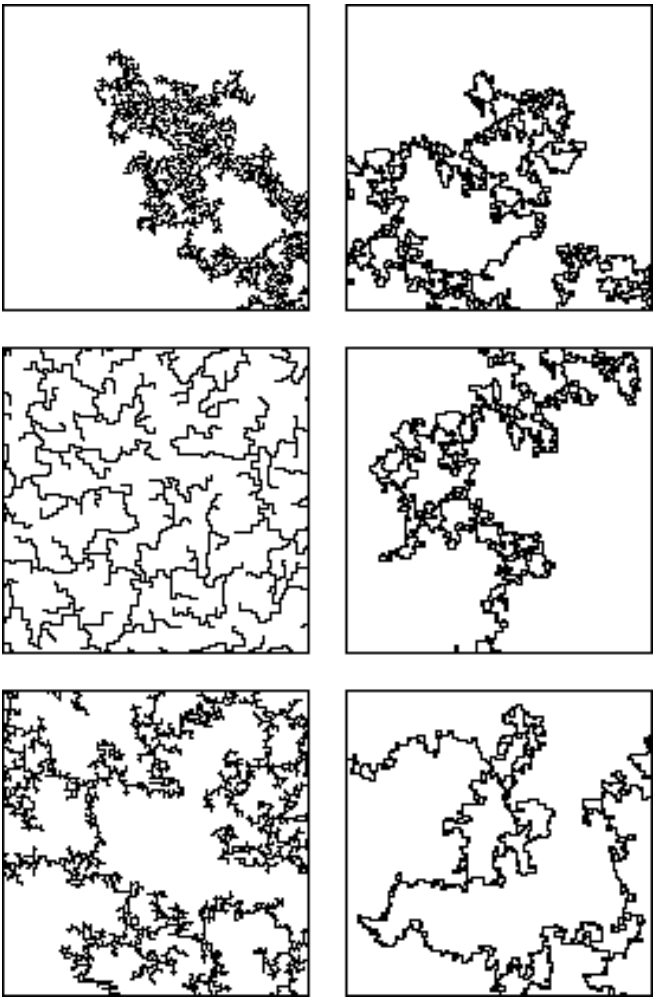}
\caption{Comparative morphology of standard site percolation
(upper line), SRSP model (middle line), and BS model (lower line).
Left column: fragments of infinite clusters at the same density
$P=0.15$. Right column: fragments of backbones at the same density
$P_B=0.1$. All fragments of size $128\times 128$. } \label{3x2}
\end{figure}

The morphology of three infinite clusters---for standard
percolation, SRSP, and BS models at the same density $P=0.15$---is
compared in Fig.\ref{3x2} (left column). The difference between
standard percolation and SRSP is striking: for percolation the
infinite cluster is strongly inhomogeneous, with loops and
dangling ends of all sizes, while for SRSP it is much more
homogeneous, practically loop-less, and apparently single-wired on
scales $r<\xi_0$, where $\xi_0$ is some nontrivial ``branching
length''. We expect $1\ll\xi_0\ll\xi_D$ at $x\ll 1$. Properties of
the infinite cluster for BS model are somewhere in between, though
closer to SRSP.

A similar comparison of three backbones at the same density
$P_B=0.1$ is made in Fig.\ref{3x2} (right column). There is no
qualitative difference between percolation and SRSP backbones,
while the BS backbone is very special: it seems to have a
``short-haired'' and almost single-wired structure: there are large loops
(of sizes $\sim \xi_B$) constituting a single-wired frame of the
backbone. The wires of this frame are dotted with very small loops
(of sizes $\sim 1$), while there are almost no loops of
intermediate sizes. This is especially well seen on
backbones of very low density ($P_B\sim 10^{-3}$, not shown).

Thus, the properties of the BS model are very different from those
of SRBP, SRSP and their hybrids: the finite density topological
phase transition, always present for all models of the SR(S/B)P
family, does not show up in BS. The reason for this effect is
apparently ``backbone reinforcement'' that accompanies sticking
events in the lattice BS models. Indeed, a finite cluster is
produced when the last bridge connecting it to the mainland is
destroyed. Upon sticking, however, not necessarily one but
possibly many new bridges are created: the larger is the
cluster, the more. These new links establish new paths that cross the
cluster and connect two opposite shores of the mainland. As a
result, connectivity of the infinite cluster increases and its
backbone strengthens.

The backbone reinforcement effect on the lattice seems to be
dramatically enhanced compared to the case of a real continual
physical system due to a special geometric resonance. A huge
number of new bonds regenerated in a single sticking event is
obviously an artifact of a lattice: it is only possible on a
lattice that many grains belonging to the cluster simultaneously
come in touch with the grains of the mainland. For
a continual system mechanically stable contact of the cluster with
the mainland would typically  be established at exactly $D$
points.

A question arises if the observed suppression of the phase
transition in the BS model is a consequence of this artificially strengthened
backbone reinforcement. Can the transition show up again in
models with more realistic moderate reinforcement effect?

\begin{figure}
\includegraphics[angle=270,width=0.9\columnwidth]{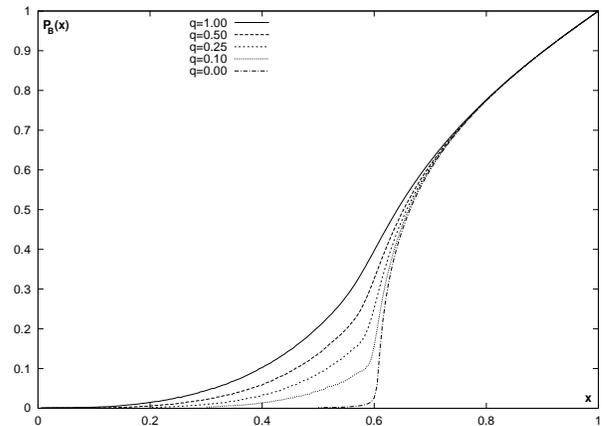}
\caption{Concentration dependence of the backbone density for a
family of BS-SRSP models with different mixing parameters $q$. The
finite density phase transition is absent  for all $q>0$. }
\label{srsp-bs-norm}
\end{figure}

\begin{figure}
\includegraphics[angle=270,width=0.9\columnwidth]{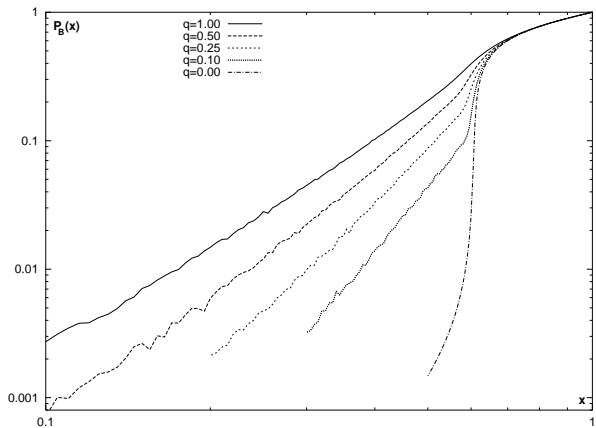}
\caption{The same as in Fig.\ref{srsp-bs-norm}, but in log-log
plot. The power-law decay \eqref{scaling2} of the backbone density
characterized by $q$-dependent exponent $\beta_B$ is clearly seen.
 }\label{srsp-bs-log}
\end{figure}

To address this question we started from a study of a family of
hybrid BS-SRSP models in which the repairing of the finite cluster
is performed according to the SRSP scenario with probability $1-q$
and according to the BS scenario with probability $q$. The
corresponding backbone densities are shown in
Figs.\ref{srsp-bs-norm},\ref{srsp-bs-log}. Although for
$x>x_c^{\rm (SRSP)}$ curves do not differ much from the SRSP
curve, for $x<x_c^{\rm (SRSP)}$ nonzero backbone density is found
for all $q>0$. The low density behavior of $P_B$ is still governed
by the power law \eqref{scaling2}, but with $q$-dependent index
$\beta_B(q)$, shown in Table \ref{table1}. Probably $\beta_B(q)$
diverges as $q\to 0$.
\begin{table}
\begin{tabular}{|c||c|c|c|c|}
\hline
$q$ & 1 & 0.5 & 0.25 & 0.1 \\
\hline
$\beta_B$  & 2.85(15) & 3.4(1)& 4.2(1) & 5.2(1)\\
\hline
\end{tabular}
\caption{The backbone density index $\beta_B$ (see
\eqref{scaling2}) for the BS-SRSP hybrid models with different
mixing parameters $q$. }\label{table1}
\end{table}

Thus, already an infinitesimal involvement of the BS process
destroys the phase transition and revives the backbone at all
nonzero concentrations. The clusters are especially large in the
vicinity of the phase transition. Therefore, one can argue that,
when the system approaches the transition, the average number of
new bridges per step sooner or later becomes large  even for rare
sticking events ($q\ll 1$),  and further destruction of the
backbone eventually slows down. However, we will see in what
follows that the large number of new bonds is not essential for
the backbone reinforcement: actually a pair of new bonds on the
opposite sides of the cluster already does the job, since it
establishes a new long path that crosses the cluster and connects
two opposite shores of the mainland.

\begin{figure}
\includegraphics[width=0.9\columnwidth]{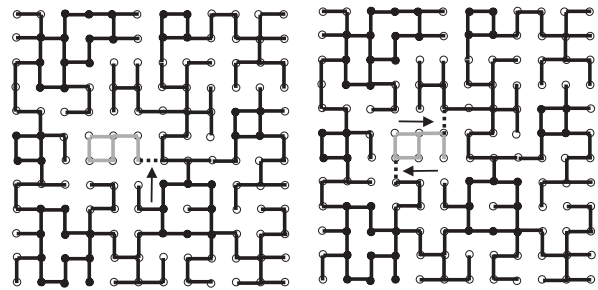}
\caption{A two-point sticking event in the SRBP(2) model. Left
panel: the dashed bond (indicated by arrow) is removed and a
finite cluster (shown grey) is created. Right panel: The finite
cluster is reconnected to the mainland by restoring two bonds
(dashed).
 }\label{srb2p}
\end{figure}

To demonstrate this we consider the simplest model with moderate
backbone reinforcement: a modified variant of SRBP with artificial
``two-point sticking''. The modification concerns only the repair
procedure which is launched after a detachment of a finite
cluster. While in the standard SRBP model it was always the
regeneration of the bond just removed, in the modified SRBP(2)
this procedure is branched:
\begin{itemize}
\item if the previous step involved the reparation procedure, then
the standard bond regeneration (as in pure SRBP) is always chosen
in the present step; \item if the previous step has lead to
removal of a bond without a violation of connectivity, then with
probability $1-k$ standard regeneration takes place, while with
probability $k$ the ``two-point sticking event'' occurs: {\it two}
randomly chosen bonds establishing contact of the cluster with the
mainland are restored (see Fig.\ref{srb2p}).
\end{itemize}

\begin{figure}
\includegraphics[angle=270,width=1\columnwidth]{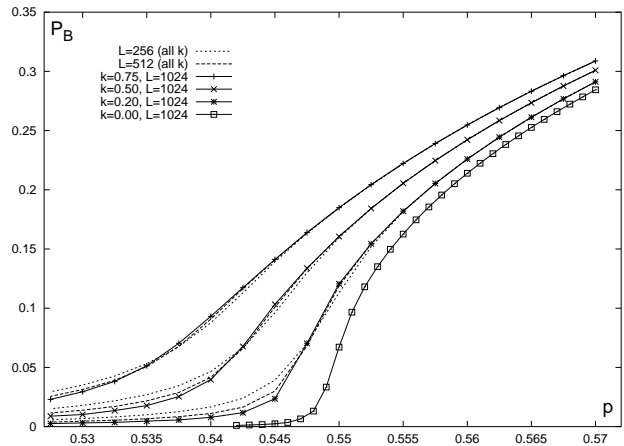}
\caption{Concentration dependence of the backbone density for the
family of modified SRBP(2) models with artificial two-point
sticking. The upper three bundles of curves correspond to three
different values of the sticking probability $k$; different curves
in each bundle were obtained for different systems sizes $L$. The
lower curve (at $k=0$) corresponds to pure SRBP model with the
phase transition.} \label{ps1}
\end{figure}

Simulation of the SRBP(2) model shows (Fig.\ref{ps1}) that the
phase transition is smeared and the backbone exists at all
accessible concentrations of bonds $p>p_{\rm tree}$ ($p_{\rm
tree}=1/2$ for the square lattice (see \cite{srbp}) for all $k>0$.
It makes one assume that the ``geometric resonance'' effect of the
lattice BS model is not essential for the suppression of the
topological phase transition, and that the BS model is not
pathological and may be trusted in this respect. It can not be
excluded that physically reasonable modifications of the BS model
exist with reinforcement effect too weak to destroy the
finite concentration phase transition. Presently we are studying
several candidates for such behavior, but, anyway, we believe that
absence of the phase transition is typical for models with
backbone reinforcement.

There is an apparent relation between the BS model and
conventional diffusion-limited cluster-cluster aggregation (DLCA)
 (see \cite{gels,jullien-botet}). Namely, these two models
can be viewed as two opposite limits of one generalized model, in
which the pore-former particles, randomly distributed in a mixture
with matrix grains (the latter having concentration $x$), are
burned at finite rate $\Gamma$. The arising free clusters move
with finite velocity $v$ (or with finite diffusion coefficient, in
the diffusional case). Then the case of relatively slow burning
(small $\Gamma$) is obviously equivalent to the BS model, while the
case of fast burning (large $\Gamma$) is described by the DLCA.
Indeed, for large $\Gamma$ in the initial (fast) stage of the
process all the pore-former grains are burned without any
considerable motion of emerging clusters. If $x<x_{\rm perc}$,
then the infinite cluster is destroyed in the course of burning:
the system is ``dissolved''. On the time-scale
$t_b\sim\Gamma^{-1}$ the pore-former is exhausted, and the burning
process practically stops, leaving behind a gas of  disconnected
clusters (in the most interesting case $x\ll 1$ the majority of
these clusters are solitary grains). Then the second (slow) stage
of the process---the aggregation---begins. It goes exactly along
the lines of the DLCA scenario (see
\cite{jullien-botet,meakin83,botet84,gimel95,hasmy96,rottereau04}).
At first small clusters stick together eventually forming large
fractal flocks (flocculation). At certain ``gelation time''
$t_g\propto v^{-1}$ the flocks become so large that they pack the
entire volume of the system and an infinite cluster arises again
in a percolation-type manner. The size of critical flocks at the
gelation point is $\xi_F\sim x^{-\nu_F}$, the index
$\nu_F=1/(D-D_F)$ being related to the fractal dimension $D_F$ of
the flocks. At the last stage of the process residual free flocks
stick to the infinite cluster and finally the system becomes a
single-cluster one. On scale $r\gg\xi_F$ the final infinite
cluster is practically homogeneous, while for $r\ll\xi_F$ it is a
fractal with the same properties, as a solitary critical flock.

It would be extremely interesting to find out if the properties of
the final single-cluster state in two limiting cases of slow and
fast burning are similar or different. We are planning to answer
this question soon.

In conclusion, we have considered burning and sticking model of a
porous material that involves sticking of detached clusters to the
mainland, and its modifications.  Such sticking normally leads to
establishing many contacts (at least two) between the cluster and
the mainland, which dramatically increases the number of
independent paths in the infinite cluster, and therefore leads to
backbone reinforcement. The latter effect is manifested in the
absence of the topological phase transition (the latter is present
in all models without reinforcement). The backbone persists up to
$x=0$, and the system, strictly speaking, remains in the net-like
phase at all $x>0$. At low $x$, however, the backbone is very
loose and feeble, its density obeys the power law \eqref{scaling2}
with non-universal exponent depending on parameters of the model.
The conductivity of the system is extremely poor at low $x$
(though finite, in contrast with the tree-like phase). The
backbone has almost single-wired structure with characteristic
size of loops $\xi_B$, which is much larger than $\xi_D$---the
characteristic spatial scale for the density correlations in the
entire system. It means that the macroscopic pores in the backbone
are filled with almost homogeneous tree-like stuff. A detailed
study of the low density state will be presented elsewhere. We
have discussed the role of parasitic ``geometric resonance'' which
presumably leads to an overestimation of the backbone
reinforcement in the BS model and have demonstrated its
irrelevance.

So, we believe that no phase transition occurs in naturally
defined models with backbone reinforcement. The simulations of the
present paper were performed for the two-dimensional square
lattice, but our preliminary results make us expect that the
behavior of three-dimensional systems is similar.

This work was supported by RFBR grant 06-02-16533.

\end{document}